\documentclass[onecolumn,preprintnumbers,eqsecnum,12pt]{revtex4}
\usepackage{amsfonts}
\usepackage{amssymb}
\usepackage{amsmath}
\usepackage{epsf}
\usepackage{graphicx}
\usepackage{dcolumn}
\usepackage{bm}

\begin{document}

\title{A note on the Hawking radiation calculated by the quasi-classical tunneling method}

\author{Ya-Peng Hu$^{a,c}$\footnote{e-mail address:
yapenghu@itp.ac.cn}, Jing-Yi Zhang$^{b,}$\footnote{e-mail address:
physicz@tom.com}, Zheng Zhao$^{d,}$\footnote{e-mail address:
zhaoz43@hotmail.com}}

\address{$^{a}$
Key Laboratory of Frontiers in Theoretical Physics, Institute of
Theoretical Physics, Chinese Academy of Sciences, P.O. Box 2735,
Beijing 100190, China}
\address{$^{b}$
Center for Astrophysics , Guangzhou University, Guangzhou 510006,
China}
\address{$^{c}$
Graduate School of the Chinese Academy of Sciences, Beijing 100039,
China}
\address{$^{d}$Department of Physics, Beijing Normal University, Beijing 100875, China}

%\vspace*{0cm}
\begin{abstract}
Since Parikh and Wilczek's tunneling method was proposed, there have
been many generalizations, such as its application to massive
charged particles' tunneling and other spacetimes. Moreover, an
invariant tunneling method was also recently proposed by Angheben et
al that it was independent of coordinates. However, there are some
subtleties in the calculation of Hawking radiation, and particularly
is the so-called factor of 2 problem during calculating the Hawking
temperature. The most popular opinion on this problem is that it is
just a problem of the choice of coordinates. However, following
other treatments we show that we can also consider this problem to
be that we do not consider the contribution from P(absorption).
Moreover, we also clarify some subtleties in the balance method and
give some comparisons with other treatments. In addition, as Parikh
and Wilczek's original works have showed that if one takes the
tunneling particles' back-reaction into account, the Hawking
radiation would be modified, and this modification is underlying
consistent with the unitary theory, we further find that this
modification is also underlying correlated with the laws of black
hole thermodynamics. Furthermore, we show that this tunneling method
may be valid just when the tunneling process is reversible.


\end{abstract}

\maketitle

\newpage

\section{Introduction}
In 2000, Parikh and Wilczek proposed a new method to reconsider the
Hawking radiation. Besides treating the Hawking radiation as a
tunneling process, they also took the tunneling particles'
back-reaction into account. And after that, they obtained the
corresponding modified spectrum. The most interesting result was
that they found this modified spectrum was implicitly consistent
with the unitary theory and could support the conservation of
information~\cite{Parikh1, Parikh2, Parikh3, Parikh4}. Following
this tunneling method, there have been many generalizations, such as
its application to other spacetimes and the massive charged
particles' tunneling cases\cite{Zhang1, Zhang2, Vagenas1, Vagenas2,
Vagenas3, Vagenas4, Vagenas5, Vagenas6, Vagenas7, Setare1, Setare2,
Setare3, Majhi1,
Majhi2,Majhi3,Majhi4,Majhi5,Majhi6,Majhi7,Majhi8,Modak,
WuandJiang1,WuandJiang2,WuandJiang3,WuandJiang4,WuandJiang5,WuandJiang6,
Hu1,Hu2,Hu3,ZhangandZhao1,ZhangandZhao2,Dynamicalspacetimes1,Dynamicalspacetimes2,Hu4,
others1,others2,others3,others4,others5,others6,others7,others8,others9,others10}.
In addition, it can also be used to calculate the black hole
temperature~\cite{Kerner} (Note that the semi-classical tunneling
method for Hawking radiation has also been separately proposed by
G.E. Volovik in~\cite{Volovik1,Volovik2}). However, Parikh and
Wilczek's tunneling method is dependent on coordinates, which means
that it should find a Painleve-like coordinates. Recently, Angheben
et al found an invariant tunneling method which was independent of
coordinates and called the Hamilton-Jacobi tunneling method to
calculate the Hawking temperature~\cite{Angheben}. This variant
tunneling method could also be considered as an extension of the
method used by Padmanabhan et
al~\cite{Complex1,Complex2,Complex3,Complex4,Complex5}.

In the above two tunneling methods, they both involve calculating
the imaginary part of the action for the (classically forbidden)
process of s-wave emission across the horizon. And according to the
WKB approximation, the tunneling probability usually related to the
imaginary part is
\begin{equation}
\Gamma \varpropto \exp (-2\text{Im}S_{out})=\exp (-2\text{Im}%
\int_{r_{in}}^{r_{out}}p_{r}dr). \label{emission1}
\end{equation}
where S is the action of the trajectory. Thus if we ignore the
tunneling particles' back-reaction effect, we can give the Hawking
temperature from
\begin{equation}
\Gamma \varpropto \exp (-2\text{Im}S_{out})=\exp (-\beta \omega ).
\label{emission2}
\end{equation}
where $\beta$ is the inverse temperature of the horizon and $\omega$
is the energy of the tunneling particle. In other words, the Hawking
temperature can be
recovered at linear order when  $2$%
Im$S_{out}=\beta \omega +o(\omega ^{2}),$ and the higher order terms
are a self-interaction effect resulting from energy
conservation~\cite{Kerner,Kraus1,Kraus2,Kraus3}. However, recently
some authors proposed that the formalism
Im$S_{out}=$Im$\int_{r_{in}}^{r_{out}}p_{r}dr$ in~(\ref{emission1})
was not invariant under canonical transformations, thus the
tunneling probability was not a proper
observable~\cite{Chowdhury,Pilling,Akhmedova1,Akhmedova2}. Moreover,
if we calculate the imaginary part of the action in different
coordinate systems by using the Hamilton-Jacobi tunneling method
and~(\ref{emission1}), we can obtain different results which means
obtaining different temperatures. For example, if we calculate the
Schwarzschild black hole's temperature in Schwarzschild static
coordinates by using the Hamilton-Jacobi tunneling method, we will
obtain the temperature $T=1/4\pi M$, which is twice the temperature
originally calculated by Hawking or other methods~\cite{Hawking,
Euclidean, D-R method, Complex1,Complex2,Complex3,Complex4,Complex5,
Anomaly1,Anomaly2,Anomaly3,Anomaly4,Anomaly5,Anomaly6,Anomaly7,Anomaly8}.
It is the so-called factor of 2 problem~\cite{Pilling,
Akhmedova1,Akhmedova2}. Note that, however, if we also
use~(\ref{emission1}) and the Hamilton-Jacobi tunneling method to
calculate the Hawking temperature in the Painleve coordinates, we
will obtain the correct Hawking temperature. Thus it can be
naturally simply considered the factor of 2 problem just as a
problem of the choice of coordinates. However, there are also other
treatments. And basically expect to generalize the tunneling method
to be independent of coordinates, Angheben et al in the original
paper of the Hamilton-Jacobi tunneling method have proposed a
treatment to solve this problem. And they argued that one had to
make a change of some spatial variable to be the corresponding
proper spatial variable defined by the spatial
metric~\cite{Angheben}. However, as Akhmedov et al found that it
couldn't in fact solve the factor of 2 problem, which could be
simply understood from the fact that the simple change of spatial
variable shouldn't change the value of the integral (more details of
argument could be seen in the references~\cite{Angheben, Pilling,
Akhmedova1,Akhmedova2}, and the key point is that the contour
associated with the divergent part of the integral is also changed
after making the radial variable to be the proper spatial variable).
Thus, following the treatment of generalizing the tunneling method
to be independent of coordinates, some authors argued that perhaps
there was a temporal imaginary contribution from the time part of
action which was usually neglected in previous works~\cite{Pilling,
Akhmedova1,Akhmedova2}. And others proposed that an integration
constant should be inserted into the action like
$S=\int_{r_{in}}^{r_{out}}p_{r}dr+C$, in which the constant C may
also give the contribution to the imaginary part of the
action~\cite{Mitra1,Mitra2}. In this note, we show that we can also
consider this problem not as a problem of the choice of coordinates
but the problem that we do not consider the contribution from P
(absorption). And we show again that if we consider the thermal
balance and take the tunneling probability such that $\Gamma
\varpropto P(emission)/P(absorption)$ (the so-called balance
method), there is no the factor of 2 problem, as have been
implicated by some previous works which recover the Hawking
temperature~\cite{Complex1,Complex2,Complex3,Complex4,Complex5}. In
addition, this tunneling probability is invariant under canonical
transformations. Furthermore, we also clarify some subtleties in the
balance method and give some comments and comparisons with other
treatments. And here the meaning of showing again is that some
authors have also considered the thermal balance before but they
gave the different result that the balance method could not solve
the factor of 2 problem~\cite{Mitra1}.

In addition, note that the back-reaction of the tunneling particles
is neglected when we recover the usual Hawking temperature. And if
one takes the back-reaction of the tunneling particles into
account~\cite{Kraus1,Kraus2,Kraus3}, the Hawking radiation would be
modified. In Parikh and Wilczek's original works they have showed
that if one takes the tunneling particles' back-reaction which comes
from the conservation of energy into account, the modified spectrum
could also be calculated by the tunneling method. And the modified
spectrum is underlying consistent with the unitary
theory~\cite{Parikh1, Parikh2, Parikh3, Parikh4}. In this note, by
taking the general Reissner-Nordstrom black hole (R-N black hole)
and kerr black hole as examples~\cite{Zhang1, Zhang2}, we could
further find that this modification is also underlying correlated
with the laws of black hole thermodynamics. Furthermore, we show
that this tunneling method may be valid only when the tunneling
process is reversible~\cite{relation}.

The outline of our paper is as follows. In Sec.~II, by taking also
the simple Schwarzschild black hole as an example, we emphasize that
the balance method could indeed solve the factor of 2 problem well.
And we also clarify some subtleties in this method and give
comparisons with other treatments. In Sec.~III, we take the R-N
black hole and kerr black hole as examples to show the underlying
correlation between the modification spectrum and the laws of black
hole thermodynamics. Section. IV. is devoted to conclusion and
discussion.

%%==========================================================

\section{Thermal balance's tunneling probability and the temperature }
The Schwarzschild black hole in the static coordinates is
\begin{equation}
ds^{2}=-(1-\frac{2M}{r})dt^{2}+(1-\frac{2M}{r})^{-1}dr^{2}+r^{2}d\Omega
^{2}.  \label{staticcoordinates}
\end{equation}
In the Hamilton-Jacobi tunneling method, the classical action $S$ of
the tunneling particle satisfies the relativistic Hamilton-Jacobi
equation~\cite{Angheben, Kerner, Pilling, Akhmedova1,Akhmedova2}
\begin{equation}
g^{\mu \nu }\partial _{\mu }S\partial _{\nu }S+m^{2}=0.
\label{Hamilton-Jacobi equation}
\end{equation}

For this metric and radial trajectories which are independent of
$\theta ,\varphi $, the Hamilton-Jacobi equation becomes
\begin{equation}
-(1-\frac{2M}{r})^{-1}(\partial _{t}S)^{2}+(1-\frac{2M}{r})(\partial
_{r}S)^{2}+m^{2}=0   \label{Hamilton-Jacobi equation2}
\end{equation}
As usual, due to the symmetries of the metric, there exists a
solution of the form
\begin{equation}
S=-\omega t+W(r)  \label{Hamilton-Jacobi ansatz}
\end{equation}%
where
\begin{equation*}
\partial _{t}S=-\omega ,\text{ }\partial _{r}S=W^{'}(r)
\end{equation*}%
Solving $W(r)$ yields \
\begin{equation}
W(r)=\pm \int \frac{dr}{(1-\frac{2M}{r})}\sqrt{\omega ^{2}-m^{2}(1-\frac{2M}{%
r})}   \label{Hamilton-Jacobi solution}
\end{equation}%
where the $+(-)$ sign in front of this integral expresses the
ingoing(outgoing) particles. Thus, it can be easily found that if we
simply consider the tunneling probability
\begin{equation}
\Gamma \varpropto \exp (-2\text{Im}S_{out})=\exp (2\text{Im}
\int_{2M-\epsilon}^{2M+\epsilon}\frac{dr}{(1-\frac{2M}{r})}\sqrt{\omega
^{2}-m^{2}(1-\frac{2M}{r})}) =\exp (-4\pi M\omega )
\label{temperature1}
\end{equation}%
where the contour lies in the upper complex plane and the minus
dropped in front of the imaginary part of the action corresponds to
the initial condition that $\partial _{r}S>0$ at $r=2M-\epsilon
<2M$~\cite{Complex1,Complex2,Complex3,Complex4,Complex5}, we will
obtain the Hawking temperature $T=1/4\pi M$. It is twice the
temperature originally calculated by Hawking and other
methods~\cite{Angheben, Kerner, Pilling, Akhmedova1,Akhmedova2}.
That is the so-called factor of 2 problem. However, if we also
use~(\ref{emission1}) and the Hamilton-Jacobi tunneling method to
calculate the Hawking temperature in the Painleve coordinates, we
will obtain the correct Hawking temperature. And the Painleve
coordinates could be obtained by the following transformation from
the Schwarzschild coordinates
\begin{equation}
t\rightarrow t_{p}-\int \frac{\sqrt{\frac{2M}{r}}}{1-2M/r}dr
\label{transformation}
\end{equation}%
and the metric in the Painleve coordinates is
\begin{equation}
ds^{2}=-(1-\frac{2M}{r})dt_{p}^{2}+2\sqrt{\frac{2M}{r}}dt_{p}dr+dr^{2}+r^{2}d\Omega
^{2}  \label{Painleve}
\end{equation}%
where one of the advantages in this coordinates is that the metric
is regular at the horizon~\cite{Parikh1, Parikh2, Parikh3, Parikh4}
(note that here the contour of the divergent part of the integral in
the Painleve coordinates is the same as that in the static
Schwarzschild coordinates, which is different from the contour in
the proper spatial variable coordinates referred
in~\cite{Angheben}). In addition, another well-known point is that
the original tunneling method first proposed by Parikh and Wilczek
should be calculated in the Painleve coordinates. Thus it can be
naturally simply considered the factor of 2 problem just as a
problem of the choice of coordinates. However, there are also other
treatments which expect to generalize the tunneling method to be
independent of coordinates. And some authors argued that when
calculating the Hawking temperature in the Schwarzschild static
coordinates perhaps there was a temporal imaginary contribution from
the time~\cite{Pilling, Akhmedova1,Akhmedova2}, while some authors
thought that one should consider the integration constant $C$ when
yields $W(r)$ in~(\ref{Hamilton-Jacobi
solution})~\cite{Mitra1,Mitra2}. Here, what we will show is mainly
that we can also consider it not to be a problem of the choice of
coordinates but to be the problem that we do not consider the
contribution from P (absorption). And we show again that if we
consider the thermal balance and take the tunneling probability as
$\Gamma \varpropto P(emission)/P(absorption)$ (the so-called balance
method), there is no the factor of 2 problem as some works have
implied~\cite{Complex1,Complex2,Complex3,Complex4,Complex5, Baocheng
Zhang}, which could be easily checked in the static Schwarzschild
coordinates as a simple example that
\begin{equation}
\Gamma \varpropto \frac{P(emission)}{P(absorption)}=\frac{\exp (-2\text{Im}%
S_{out})}{\exp (-2\text{Im}S_{in})}=\frac{\exp (-4\pi M\omega
)}{\exp (4\pi M\omega )}=\exp (-8\pi M\omega ) \label{temperature2}
\end{equation}
which can recover the Hawking temperature $T=1/8\pi M$. It should be
noted that although the above treatment has already been proposed by
T.Padmanabhan et al, there are some subtleties in this method. And
these subtleties may be the direct motivation of the other two
treatments. The first subtlety is that $P(absorption)$
in~(\ref{temperature2}) is greater than unity. Thus, if it is
considered as the absorption probability, it would be unphysical.
However, considering $P(absorption)$ as the absorption probability
is not suitable here because we have neglected the unitary factor in
front of the wave function. And it can be easily checked out that
considering $P(emission)/P(absorption)$ as the tunneling probability
is always suitable. The second subtlety is that the tunneling
probability $\Gamma \varpropto P(emission)/P(absorption)$ is indeed
invariant under canonical transformations, which means that the
tunneling probability has the same result in any coordinates. The
third subtlety is that whether the time in~(\ref{Hamilton-Jacobi
ansatz}) can give the contribution to the imaginary part of the
action. In the present status, we think that it is not necessary to
consider its contribution. The main reason is just because the
temporal imaginary contribution from the time can be canceled
automatically in P(emission)/P(absorption). In fact, after some
simple calculations, we could find that during the calculation in
the Schwarzschild static coordinates the contribution from the $\exp
(-2$Im$S_{in})$ in~(\ref{temperature2}) is just equal to the
contribution from the time in references~\cite{Pilling,
Akhmedova1,Akhmedova2} or the integration constant in
references~\cite{Mitra1,Mitra2}. Moreover, it should also be easily
found that the contribution from the $\exp (-2$Im$S_{in})$ is just
equal to unit in the Painleve coordinates (which is correlated with
the order that the metric in Painleve-like coordinates should be
regular at the horizon, then $S_{in}$ is real), thus the tunneling
probability is $\Gamma \varpropto \exp (-2$Im$S_{out})$ which is
just the familiar formalism~(\ref{emission1}) in the original
tunneling method's treatment~\cite{Parikh1,Parikh2,Parikh3,Parikh4}.
Thus it can also be considered as another underlying reason why it
should find a Painleve-like coordinates and the tunneling
probability in~(\ref{emission1}) can work well in the original
tunneling method.

\section{The modified Hawking radiation and its relationship with laws of black hole thermodynamics}
In the above section, we calculate the Hawking temperature by
dropping effect of the back-reaction of the tunneling particle. In
this section, we take it into
account~\cite{Parikh1,Parikh2,Parikh3,Parikh4,
Kraus1,Kraus2,Kraus3}. In Parikh and Wilczek's original works they
have shown that the back-reaction effect will give a correction to
the Hawking radiation (i.e, modified spectrum), and this
modification is underlying consistent with the unitary theory and
can support the conservation of information during the evolution of
black hole~\cite{Parikh1,Parikh2,Parikh3,Parikh4}. In addition, it
has also been found that this modification is underlying correlated
with the laws of thermodynamics. Because although the behavior of
massive particles' tunneling is different from that of massless
particles' tunneling, two integrations are the same after we first
integrate the radial part during the calculation of the imaginary
part of the action, and the integration is direct proportional to
the inverse of the temperature. Moreover, this integration can be
expressed as the formalism of laws of thermodynamics in some
cases~\cite{Hu1,Hu2,Hu3,ZhangandZhao1,ZhangandZhao2,relation}. In
the following, taking the more general R-N black hole and Kerr black
hole as examples, we further show that the underlying relationship
between the modification and laws of thermodynamics is universal.
Furthermore, we show that this tunneling method may be valid only
when the tunneling process is reversible.

\subsection{The R-N black hole}

Recently, Parikh and Wilczek's original works have been extended to
the R-N black hole. As a general solution with a simple charge in
Einstein equation, we first take it as an example. According to
Ref~\cite{Zhang2}, after taking the back-reaction of the tunneling
massive charged particle into account, the imaginary part of the
action for the classically forbidden trajectory is
\begin{eqnarray}
\text{Im}S &=&\text{Im}\{\int\nolimits_{r_{i}}^{r_{f}}[p_{r}-\frac{p_{A_{t}}%
\overset{\cdot }{A_{t}}}{\overset{\cdot }{r}}]dr\}  \notag \\
&=&-\text{Im}\{\int\nolimits_{r_{i}}^{r_{f}}\int_{(M,Q)}^{(M-\omega ,Q-q)}[%
\frac{2r\sqrt{2Mr-Q^{2}}}{r^{2}-2Mr+Q^{2}}dM-\frac{2\sqrt{2Mr-Q^{2}}Q}{%
r^{2}-2Mr+Q^{2}}dQ]dr\}  \notag \\
&=&-\pi \int_{(M,Q)}^{(M-\omega ,Q-q)}[\frac{(M+\sqrt{M^{2}-Q^{2}})^{2}}{%
\sqrt{M^{2}-Q^{2}}}dM-\frac{(M+\sqrt{M^{2}-Q^{2}})Q}{\sqrt{M^{2}-Q^{2}}}dQ]
\notag \\
&=&-\frac{\pi }{2}\{[(M-\omega )+\sqrt{(M-\omega )^{2}-(Q-q)^{2}}]^{2}-[M+%
\sqrt{M^{2}-Q^{2}}]^{2}\}  \notag \\
&=&-\frac{1}{2}\Delta S_{BH}.  \label{R-N black hole}
\end{eqnarray}%
where $A_{t}=\frac{Q}{r}$ is the first component of the
4-Dimensional electromagnetic potential, and $p_{A_{t}}$ is the
corresponding canonical momentum conjugate. As the conclusion
discussed in Ref~\cite{Zhang2}, from ~(\ref{R-N black hole}) we can
easily obtain the modified spectrum and find that the original
temperature of R-N black hole is just the case of neglecting the
higher-order terms of $\omega$ and $q$. Moreover, it can also be
found that this modified spectrum is consistent with the underlying
unitary theory, which supports the conservation of information.
Here, we further show that an interesting result could also be
obtained in the same equation~(\ref{R-N black hole}) if we start
from the viewpoint of laws of black hole thermodynamics, which is
expressed in the following. And the underlying relationship between
the tunneling process and the laws of black hole thermodynamics may
give a new insight into the tunneling process.

As we know, for the R-N black hole, when a charged massive particle
tunnels across the event horizon, the mass and the charge of black
hole will be changed as a consequence. According to the first law of
black hole thermodynamics, the differential Bekenstein-Smarr
equation of the R-N black hole is~\cite{Bekenstein1,Bekenstein2}
\begin{equation}
dM=\frac{\kappa }{8\pi }dA+VdQ\text{ }(J=0),  \label{B-S equation1}
\end{equation}%
Furthermore, if the tunneling process is considered as a reversible
process, according to the second law of black hole thermodynamics,
~(\ref{B-S equation1}) can be rewritten as
\begin{equation}
dM=TdS+VdQ.  \label{reversible equation1}
\end{equation}%
Equally, it can be rewritten as
\begin{equation}
dS=\frac{dM}{T}-\frac{VdQ}{T}.  \label{reversible equation2}
\end{equation}
The temperature and the potential are respectively~\cite{Zhang2}
\begin{equation}
T=\frac{\sqrt{M^{2}-Q^{2}}}{2\pi (M+\sqrt{M^{2}-Q^{2}})^{2}},V=\frac{Q}{M+%
\sqrt{M^{2}-Q^{2}}}.  \label{temperature and the potential}
\end{equation}%
Substituting~(\ref{temperature and the potential})
into~(\ref{reversible equation2}), we can obtain
\begin{equation}
dS=\frac{2\pi
(M+\sqrt{M^{2}-Q^{2}})^{2}}{\sqrt{M^{2}-Q^{2}}}dM-\frac{2\pi
(M+\sqrt{M^{2}-Q^{2}})Q}{\sqrt{M^{2}-Q^{2}}}dQ.  \label{reversible
equation3}
\end{equation}%
Thus, using~(\ref{reversible equation3}) we can rewrite~(\ref{R-N
black hole}) as
\begin{eqnarray}
\text{Im}S &=&-\pi \int_{(M,Q)}^{(M-\omega ,Q-q)}[\frac{(M+\sqrt{M^{2}-Q^{2}}%
)^{2}}{\sqrt{M^{2}-Q^{2}}}dM-\frac{(M+\sqrt{M^{2}-Q^{2}})Q}{\sqrt{M^{2}-Q^{2}%
}}dQ]  \notag \\
&=&-\frac{1}{2}\int_{S_{i}}^{S_{f}}dS=-\frac{1}{2}\Delta S_{BH}.
\label{rewriteEq}
\end{eqnarray}%
Which is also the same result as~(\ref{R-N black hole}). The
difference is that the result in ~(\ref{rewriteEq}) implicates that
Hawking radiation via tunneling is correlated with the laws of black
hole thermodynamics.

\subsection{The Kerr black hole}
As another general solution, the Kerr black hole spacetime with a
simple angular momentum, we also take it as an example to show the
relationship between the modified spectrum and the laws of black
hole thermodynamics. For the sake of simplicity, we only investigate
the case of a massless particle with angular momentum tunneling
across the outer event horizon.

According to Ref~\cite{Zhang1}, after taking the tunneling massless
particles' back-reaction into account, the imaginary part of the
action for the classically forbidden trajectory is
\begin{eqnarray}
\text{Im}S &=&\text{Im}[\int_{r_{i}}^{r_{f}}p_{r}dr-\int_{\varphi
_{i}}^{\varphi _{f}}p_{\varphi }d\varphi ]  \notag \\
&=&\text{Im}[\int\nolimits_{r_{i}}^{r_{f}}\int_{M}^{M-\omega }\frac{\sqrt{%
(r^{2}+a^{2})^{2}-\Delta a^{2}\sin ^{2}\theta }}{\rho ^{2}-\sqrt{\rho
^{2}(\rho ^{2}-\Delta )}}drdM  \notag \\
&&-\int\nolimits_{r_{i}}^{r_{f}}\int_{M}^{M-\omega }\frac{\sqrt{%
(r^{2}+a^{2})^{2}-\Delta a^{2}\sin ^{2}\theta }}{\rho ^{2}-\sqrt{\rho
^{2}(\rho ^{2}-\Delta )}}a\Omega drdM]  \notag \\
&=&\int_{M}^{M-\omega }\frac{-2\pi (M^{2}+M\sqrt{M^{2}-a^{2}})}{\sqrt{%
M^{2}-a^{2}}}dM+\int_{M}^{M-\omega }\frac{\pi a^{2}}{\sqrt{M^{2}-a^{2}}}dM
\notag \\
&=&\pi \lbrack M^{2}-(M-\omega )^{2}+M\sqrt{M^{2}-a^{2}}-(M-\omega )\sqrt{%
(M-\omega )^{2}-a^{2}}  \notag \\
&=&-\frac{1}{2}\Delta S_{BH}.  \label{Kerrbh}
\end{eqnarray}

As the same discussion as that of R-N black hole, we next rewrite
~(\ref{Kerrbh}) from the viewpoint of laws of black hole
thermodynamics. For the Kerr black hole, the first law of black hole
thermodynamics is~\cite{Bekenstein1,Bekenstein2}
\begin{equation}
dM=\frac{\kappa }{8\pi }dA+\Omega dJ\text{ }(Q=0),  \label{B-S
equation2}
\end{equation}
And if the tunneling process is a reversible process, we can obtain
\begin{equation}
dS=\frac{dM}{T}-\frac{\Omega dJ}{T}.  \label{reversibleEqKerr1}
\end{equation}
The temperature, the angle velocity and the angular momentum of Kerr
black hole are respectively~\cite{Zhang1}
\begin{equation}
T=\frac{\sqrt{M^{2}-a^{2}}}{4\pi (M^{2}+M\sqrt{M^{2}-a^{2}})},\Omega =\frac{a%
}{r_{+}^{2}+a^{2}}=\frac{a}{2(M^{2}+M\sqrt{M^{2}-a^{2}})},J=aM.
\label{temperature and angle velocity}
\end{equation}%
Thus, substituting~(\ref{temperature and angle velocity})
into~(\ref{reversibleEqKerr1}), we get
\begin{equation}
dS=\frac{4\pi
(M^{2}+M\sqrt{M^{2}-a^{2}})}{\sqrt{M^{2}-a^{2}}}dM-\frac{2\pi
a^{2}}{\sqrt{M^{2}-a^{2}}}dM.  \label{reversibleEqKerr2}
\end{equation}%
After comparing~(\ref{reversibleEqKerr2}) with~(\ref{Kerrbh}), it is
easy to find that we can also rewrite the imaginary part of the
action as follows
\begin{eqnarray}
\text{Im}S &=&\int_{M}^{M-\omega }\frac{-2\pi (M^{2}+M\sqrt{M^{2}-a^{2}})}{%
\sqrt{M^{2}-a^{2}}}dM+\int_{M}^{M-\omega }\frac{\pi a^{2}}{\sqrt{M^{2}-a^{2}}%
}dM  \notag \\
&=&-\frac{1}{2}\int_{M}^{M-\omega }[\frac{4\pi (M^{2}+M\sqrt{M^{2}-a^{2}})}{%
\sqrt{M^{2}-a^{2}}}dM-\frac{2\pi a^{2}}{\sqrt{M^{2}-a^{2}}}dM]  \notag \\
&=&-\frac{1}{2}\int_{S_{i}}^{S_{f}}dS=-\frac{1}{2}\Delta S_{BH}.
\label{rewriteEqKerr}
\end{eqnarray}%
which shows again that the Hawking radiation via tunneling is also
correlated with the laws of black hole thermodynamics for the
stationary axial symmetry case.

\section{Conclusion and Discussion}
In this paper, we mainly give a note and make a discussion on the
Hawking radiation calculated by the quasi-local tunneling method.
And we show that the original Hawking temperature will be recovered
if we neglect the tunneling particles' back-reaction effect. And if
we take the effect into account, the original Hawking radiation
would be modified.

During the calculation of original Hawking temperature, following
the treatments which do not simply consider the factor of 2 problem
just as a problem of choice of coordinates, we show again that the
so-called factor of 2 problem could indeed be solved well by
considering the thermal balance and taking the tunneling probability
as $\Gamma \varpropto P(emission)/P(absorption)$. Moreover, we also
clarify some subtleties in this balance method and compare this
treatment with other treatments. In addition, we also find out why
the familiar formalism~(\ref{emission1}) works well in the original
tunneling method proposed by Parikh and Wilczek. Because the
underlying advantage of Painleve-like coordinates can make the
$P(absorption)$ to be unit. In spite of that, here we would also
give some comments on other treatments. In the constant C
treatment~\cite{Mitra1,Mitra2}, it may be a little unphysical since
one essentially just picks C to be whatever is needed to get the
original Hawking result. Thus there is no physical content in this
approach. More mathematically notice that S is a definite integral
(i.e. you have limits $r_{in}$ and $r_{out}$) thus there could be no
integration constant. In addition, note that a direct motivation of
this treatment is that because $P(absorption)$ may be greater than
unity. However, as we have discussed in Sec II, this is a subtlety
in the balance method. And it is just because we neglect the unitary
factor in front of the wave function. While in the treatment
considering the temporal imaginary contribution from the time,
although there are some works based on it~\cite{Majhi1,
Majhi2,Majhi3,Majhi4,Majhi5,Majhi6,Majhi7,Majhi8,Modak, Pilling,
Akhmedova1,Akhmedova2}, in fact, after we consider the thermal
balance, the temporal imaginary contribution from the time could be
considered as a redundance because it can be canceled automatically
in P(emission)/P(absorption). In other words, if we do not consider
the temporal imaginary contribution from the time, we can also have
the same results such as the modified temperature in some previous
works~\cite{Majhi1,
Majhi2,Majhi3,Majhi4,Majhi5,Majhi6,Majhi7,Majhi8,Modak}. In
addition, note also that the transformation in
(~\ref{transformation}) can not be considered as the evidence that
the Schwarzschild time has the temporal contribution. And it just
shows how to get the non-singular coordinates defined in the whole
region from two unconnected patches which have the same coordinate
system. Therefore, in the present status, it is not necessary to
consider the temporal contribution. And in fact the contributions
from the time (or the constant) is just equal to the contribution
from $\exp (-2$Im$S_{in})$ in the balance method's treatment. Note
that, another interesting result may be also obtained after we
considering $P(absorption)$ in the tunneling probability. It is that
the ratio between P(emission) and P(absorption) can be just the
relative scattering amplitude viewed from the Damour-Ruffini
method~\cite{D-R method}.  Thus the balance treatment may be
underlying consistent with the Damour-Ruffini method. In addition,
the Damour-Ruffini method is underlying related with two different
vacua~\cite{Sannan}. Therefore, it would be interesting to give a
further light on the balance treatment viewed from the
Damour-Ruffini method and two non-equivalent vacua. In addition,
note also that the thermodynamics on dynamical spacetime is still an
open question until now. As we know, some methods calculating the
Hawking temperature may be invalid in the dynamical spacetimes such
as the Euclidean method or anomaly method~\cite{Euclidean,
Anomaly1,Anomaly2,Anomaly3,Anomaly4,Anomaly5,Anomaly6,Anomaly7,Anomaly8},
thus the tunneling method may be a good tool to investigate the
Hawking temperature of the dynamical
spacetme~\cite{Dynamicalspacetimes1,Dynamicalspacetimes2}. And more
details about the research on the thermodynamics of dynamical
spacetimes could be seen
in~\cite{DTherml1,DTherml2,DTherml3,DTherml4,DTherml5,DTherml6,DTherml7,DTherml8,
DTherml9,DTherml10,DTherml11,DTherml12,DTherml13,DTherml14,DTherml15,DTherml16,
DTherml17,DTherml18,DTherml19,DTherml20,DTherml21,DTherml22,DTherml23,DTherml24,DTherml25,DTherml26}.

On the other hand, during the calculation of the modified Hawking
radiation, we mainly focus on researching the underlying contents.
As Parikh and Wilczek's original works have showed that this
modification is underlying consistent with the unitary theory and
can support the conservation of information, we further show that
this modification is also underlying correlated with the laws of
black hole thermodynamics by taking the general R-N black hole and
Kerr black hole for examples. And this new relationship may give a
new insight into the tunneling process. As a simple consequence, it
may imply that the tunneling method is valid only when the tunneling
process is reversible. And if the tunneling process is irreversible,
the conservation of information may be violated~\cite{relation}.
However, how to consider or add the non-equilibrium effects is also
an open question and deserves further investigations.

\section{Acknowledgements}
Ya-Peng Hu thanks Professors Rong-Gen Cai and Douglas Singleton, Dr.
Jia-Rui Sun and Xian Gao for their useful discussions. And Ya-Peng
Hu is supported partially by grants from NSFC, China (No. 10875018
and No. 10975168), and a grant from the Chinese Academy of Sciences.
Jing-Yi Zhang is supported partly by the National Natural Science
Foundation of China (Grant Nos. 10573005, 10633010 and 10873003),
the National Basic Research Program of China (Grant No.
2007CB815405), and the Natural Science Foundation of Guangdong
Province (Grant No. 7301224). Zheng Zhao is supported by the
National Natural Science Foundation of China (Grant No. 10773002)
and the National Basic Research Program of China (Grant No.
2003CB716302).


\begin{thebibliography}{99}
\bibitem{Parikh1}
  M.~K.~Parikh and F.~Wilczek,
  %``Hawking radiation as tunneling,''
  Phys.\ Rev.\ Lett.\  {\bf 85}, 5042 (2000)
  [arXiv:hep-th/9907001];
  %%CITATION = PRLTA,85,5042;%
  %\cite{Parikh:2002qh}

\bibitem{Parikh2}
  M.~K.~Parikh,
  %``New coordinates for de Sitter space and de Sitter radiation,''
  Phys.\ Lett.\  B {\bf 546}, 189 (2002)
  [arXiv:hep-th/0204107];
  %%CITATION = PHLTA,B546,189;%%
%\cite{Parikh:2004ih}
%\cite{Parikh:2004rh}

\bibitem{Parikh3}
  M.~K.~Parikh,
  %``Energy conservation and Hawking radiation,''
  arXiv:hep-th/0402166;
  %%CITATION = HEP-TH/0402166;%%

\bibitem{Parikh4}
  M.~K.~Parikh,
  %``A secret tunnel through the horizon,''
  Int.\ J.\ Mod.\ Phys.\  D {\bf 13}, 2351 (2004)
  [Gen.\ Rel.\ Grav.\  {\bf 36}, 2419 (2004)]
  [arXiv:hep-th/0405160].
  %%CITATION = GRGVA,36,2419;%%

%\cite{Zhang:2005wn}
\bibitem{Zhang1}
  J.~Y.~Zhang and Z.~Zhao,
  %``Hawking Radiation Via Tunneling From Kerr Black Holes,''
  Mod.\ Phys.\ Lett.\  A {\bf 20}, 1673 (2005).
  %%CITATION = MPLAE,A20,1673;%%

%\cite{Zhang:2005xt}
\bibitem{Zhang2}
  J.~Y.~Zhang and Z.~Zhao,
  %``Hawking radiation of charged particles via tunneling from the
  %Reissner-Nordstroem black hole,''
  JHEP {\bf 0510}, 055 (2005).
  %%CITATION = JHEPA,0510,055;%%

%\cite{Vagenas:2000am}
\bibitem{Vagenas1}
  E.~C.~Vagenas,
  %``Are extremal 2-D black holes really frozen?,''
  Phys.\ Lett.\  B {\bf 503}, 399 (2001)
  [arXiv:hep-th/0012134];
  %%CITATION = PHLTA,B503,399;%%
  %\cite{Vagenas:2001sm}

\bibitem{Vagenas2}
  E.~C.~Vagenas,
  %``BTZ black holes and Hawking radiation,''
  Mod.\ Phys.\ Lett.\  A {\bf 17}, 609 (2002)
  [arXiv:hep-th/0108147];
  %%CITATION = MPLAE,A17,609;%%
%\cite{Vagenas:2001rm}

\bibitem{Vagenas3}
  E.~C.~Vagenas,
  %``Quantum corrections to the Bekenstein-Hawking entropy of the BTZ black
  %hole via self-gravitation,''
  Phys.\ Lett.\  B {\bf 533}, 302 (2002)
  [arXiv:hep-th/0109108];
  %%CITATION = PHLTA,B533,302;%%
%\cite{Vagenas:2002hs}

\bibitem{Vagenas4}
  E.~C.~Vagenas,
  %``Generalization of the KKW analysis for black hole radiation,''
  Phys.\ Lett.\  B {\bf 559}, 65 (2003)
  [arXiv:hep-th/0209185];
  %%CITATION = PHLTA,B559,65;%%
%\cite{Medved:2005yf}

\bibitem{Vagenas5}
  A.~J.~M.~Medved and E.~C.~Vagenas,
  %``On Hawking radiation as tunneling with back-reaction,''
  Mod.\ Phys.\ Lett.\  A {\bf 20}, 2449 (2005)
  [arXiv:gr-qc/0504113];
  %%CITATION = MPLAE,A20,2449;%%
%\cite{Medved:2005vw}

\bibitem{Vagenas6}
  A.~J.~M.~Medved and E.~C.~Vagenas,
  %``On Hawking radiation as tunneling with logarithmic corrections,''
  Mod.\ Phys.\ Lett.\  A {\bf 20}, 1723 (2005)
  [arXiv:gr-qc/0505015];
  %%CITATION = MPLAE,A20,1723;%%
%\cite{Arzano:2005rs}

\bibitem{Vagenas7}
  M.~Arzano, A.~J.~M.~Medved and E.~C.~Vagenas,
  %``Hawking radiation as tunneling through the quantum horizon,''
  JHEP {\bf 0509}, 037 (2005)
  [arXiv:hep-th/0505266].
  %%CITATION = JHEPA,0509,037;%%

%\cite{Setare:2003vs}
\bibitem{Setare1}
  M.~R.~Setare and E.~C.~Vagenas,
  %``Self-gravitational corrections to the Cardy-Verlinde formula of
  %Achucarro-Ortiz black hole,''
  Phys.\ Lett.\  B {\bf 584}, 127 (2004)
  [arXiv:hep-th/0309092];
  %%CITATION = PHLTA,B584,127;%%
  %\cite{Setare:2004us}

\bibitem{Setare2}
  M.~R.~Setare and E.~C.~Vagenas,
  %``Self-gravitational corrections to the Cardy-Verlinde formula and the  FRW
  %brane cosmology in SdS(5) bulk,''
  Int.\ J.\ Mod.\ Phys.\  A {\bf 20}, 7219 (2005)
  [arXiv:hep-th/0405186];
  %%CITATION = IMPAE,A20,7219;%%
  %\cite{Setare:2008zg}

\bibitem{Setare3}
  M.~R.~Setare,
  %``Semiclassical Corrections to the Cardy-Verlinde Formula of Kerr Black
  %Holes,''
  Int.\ J.\ Mod.\ Phys.\  A {\bf 23}, 2047 (2008)
  [arXiv:0807.0273 [hep-th]].
  %%CITATION = IMPAE,A23,2047;%%



%\cite{Banerjee:2008ry}
\bibitem{Majhi1}
  R.~Banerjee and B.~R.~Majhi,
  %``Quantum Tunneling and Back Reaction,''
  Phys.\ Lett.\  B {\bf 662}, 62 (2008)
  [arXiv:0801.0200 [hep-th]];
  %%CITATION = PHLTA,B662,62;%%
%\cite{Banerjee:2008gc}

\bibitem{Majhi2}
  R.~Banerjee, B.~R.~Majhi and S.~Samanta,
  %``Noncommutative Black Hole Thermodynamics,''
  Phys.\ Rev.\  D {\bf 77}, 124035 (2008)
  [arXiv:0801.3583 [hep-th]];
  %%CITATION = PHRVA,D77,124035;%%
%\cite{Banerjee:2008cf}

\bibitem{Majhi3}
  R.~Banerjee and B.~R.~Majhi,
  %``Quantum Tunneling Beyond Semiclassical Approximation,''
  JHEP {\bf 0806}, 095 (2008)
  [arXiv:0805.2220 [hep-th]];
  %%CITATION = JHEPA,0806,095;%%
%\cite{Majhi:2009uk}

\bibitem{Majhi4}
  B.~R.~Majhi and S.~Samanta,
  %``Hawking Radiation due to Photon and Gravitino Tunneling,''
  arXiv:0901.2258 [hep-th];
  %%CITATION = ARXIV:0901.2258;%%
  %\cite{Modak:2008tg}
%\cite{Banerjee:2008fz}

\bibitem{Majhi5}
  R.~Banerjee and B.~R.~Majhi,
  %``Quantum Tunneling, Trace Anomaly and Effective Metric,''
  arXiv:0808.3688 [hep-th];
  %%CITATION = ARXIV:0808.3688;%%
  %\cite{Majhi:2008gi}

\bibitem{Majhi6}
  B.~R.~Majhi,
  %``Fermion Tunneling Beyond Semiclassical Approximation,''
  arXiv:0809.1508 [hep-th];
  %%CITATION = ARXIV:0809.1508;%%
  %\cite{Banerjee:2008sn}

\bibitem{Majhi7}
  R.~Banerjee and B.~R.~Majhi,
  %``Connecting anomaly and tunneling methods for Hawking effect through
  %chirality,''
  arXiv:0812.0497 [hep-th];
  %%CITATION = ARXIV:0812.0497;%%
  %\cite{Banerjee:2009wa}

\bibitem{Majhi8}
  R.~Banerjee, B.~R.~Majhi and D.~Roy,
  %``Corrections to Unruh effect in tunneling formalism and mapping with Hawking
  %effect,''
  arXiv:0901.0466 [hep-th];
  %%CITATION = ARXIV:0901.0466;%%

\bibitem{Modak}
  S.~K.~Modak,
  %``Corrected entropy of BTZ black hole in tunneling approach,''
  Phys.\ Lett.\  B {\bf 671}, 167 (2009)
  [arXiv:0807.0959 [hep-th]].
  %%CITATION = PHLTA,B671,167;%%

%\cite{Jiang:2005xb}
\bibitem{WuandJiang1}
  Q.~Q.~Jiang and S.~Q.~Wu,
  %``Hawking radiation of charged particles as tunneling from
  %Reissner-Nordstroem-de Sitter black holes with a global monopole,''
  Phys.\ Lett.\  B {\bf 635}, 151 (2006)
  [Erratum-ibid.\  {\bf 639}, 684 (2006)]
  [arXiv:hep-th/0511123];
  %%CITATION = PHLTA,B635,151;%%
%\cite{Jiang:2005ba}

\bibitem{WuandJiang2}
  Q.~Q.~Jiang, S.~Q.~Wu and X.~Cai,
  %``Hawking radiation as tunneling from the Kerr and Kerr-Newman black
  %holes,''
  Phys.\ Rev.\  D {\bf 73}, 064003 (2006)
  [Erratum-ibid.\  D {\bf 73}, 069902 (2006)]
  [arXiv:hep-th/0512351];
  %%CITATION = PHRVA,D73,064003;%%
%\cite{Wu:2006pz}

\bibitem{WuandJiang3}
  S.~Q.~Wu and Q.~Q.~Jiang,
  %``Remarks on Hawking radiation as tunneling from the BTZ black holes,''
  JHEP {\bf 0603}, 079 (2006)
  [arXiv:hep-th/0602033];
  %%CITATION = JHEPA,0603,079;%%
%\cite{Jiang:2007tn}

\bibitem{WuandJiang4}
  Q.~Q.~Jiang, S.~Q.~Wu and S.~Z.~Yang,
  %``New form of the Kerr-Newman-Kasuya solution and its Hawking radiation via
  %tunneling,''
  Int.\ J.\ Mod.\ Phys.\  A {\bf 22} (2007) 777;
  %%CITATION = IMPAE,A22,777;%%
%\cite{Chen:2008vi}

\bibitem{WuandJiang5}
  D.~Y.~Chen, Q.~Q.~Jiang, S.~Z.~Yang and X.~T.~Zu,
  %``Fermions tunnelling from the charged dilatonic black holes,''
  Class.\ Quant.\ Grav.\  {\bf 25}, 205022 (2008)
  [arXiv:0803.3248 [hep-th]];
  %%CITATION = CQGRD,25,205022;%%
%\cite{Jiang:2008gq}

\bibitem{WuandJiang6}
  Q.~Q.~Jiang,
  %``Dirac particles' tunnelling from black rings,''
  Phys.\ Rev.\  D {\bf 78}, 044009 (2008)
  [arXiv:0807.1358 [hep-th]].
  %%CITATION = PHRVA,D78,044009;%%

%\cite{Hu:2006tj}
\bibitem{Hu1}
  Y.~P.~Hu, J.~Y.~Zhang and Z.~Zhao,
  %``Massive particles' Hawking radiation via tunneling from the G.H Dilaton
  %black hole,''
  Mod.\ Phys.\ Lett.\  A {\bf 21}, 2143 (2006)
  [arXiv:gr-qc/0611026];
  %%CITATION = MPLAE,A21,2143;%%
%\cite{Hu:2006xv}

\bibitem{Hu2}
  Y.~P.~Hu, J.~Y.~Zhang and Z.~Zhao,
  %``Massive uncharged and charged particles' tunneling from the
  %Horowitz-Strominger dilaton black hole,''
  Int.\ J.\ Mod.\ Phys.\  D {\bf 16}, 847 (2007)
  [arXiv:gr-qc/0611085];
  %%CITATION = IMPAE,D16,847;%%
%\cite{Zhang:2005gja}

\bibitem{Hu3}
  Y.~P.~Hu, L.~Gao, and Z.~Zhao,
  %``Massive Particles' Tunneling Effect from An Arbitrarily Dimensional Schwarzschild Black Hole,''
  Int.\ J.\ Theor.\ Phys.\  {\bf 45}, 2001 (2006).
  %%CITATION = IJTPB,45,2001;%%

\bibitem{ZhangandZhao1}
  J.~Y.~Zhang and Z.~Zhao,
  %``New coordinates for Kerr-Newman black hole radiation,''
  Phys.\ Lett.\  B {\bf 618}, 14 (2005);
  %%CITATION = PHLTA,B618,14;%%
%\cite{Zhang:2005sf}

\bibitem{ZhangandZhao2}
  J.~Y.~Zhang and Z.~Zhao,
  %``Massive particles-prime black hole tunneling and de Sitter tunneling,''
  Nucl.\ Phys.\  B {\bf 725}, 173 (2005);
  %%CITATION = NUPHA,B725,173;%%
%\cite{Zhang:2005mn}

\bibitem{Dynamicalspacetimes1}
  R.~Di Criscienzo, M.~Nadalini, L.~Vanzo, S.~Zerbini and G.~Zoccatelli,
  %``On the Hawking radiation as tunneling for a class of dynamical black
  %holes,''
  Phys.\ Lett.\  B {\bf 657}, 107 (2007)
  [arXiv:0707.4425 [hep-th]];
  %%CITATION = PHLTA,B657,107;%%
%\cite{Hayward:2008jq}

\bibitem{Dynamicalspacetimes2}
  S.~A.~Hayward, R.~Di Criscienzo, L.~Vanzo, M.~Nadalini and S.~Zerbini,
  %``Local Hawking temperature for dynamical black holes,''
  arXiv:0806.0014 [gr-qc];
  %%CITATION = ARXIV:0806.0014;%%

%\cite{Cai:2008gw}
\bibitem{Hu4}
  R.~G.~Cai, L.~M.~Cao and Y.~P.~Hu,
  %``Hawking Radiation of Apparent Horizon in a FRW Universe,''
  Class.\ Quant.\ Grav.\  {\bf 26}, 155018 (2009)
  [arXiv:0809.1554 [hep-th]].
  %%CITATION = CQGRD,26,155018;%%


\bibitem{others1}
  S.~Hemming and E.~Keski-Vakkuri,
  %``Hawking radiation from AdS black holes,''
  Phys.\ Rev.\  D {\bf 64}, 044006 (2001)
  [arXiv:gr-qc/0005115];
  %%CITATION = PHRVA,D64,044006;%%

\bibitem{others2}
  K.~Nozari and S.~H.~Mehdipour,
  %``Hawking Radiation as Quantum Tunneling from Noncommutative Schwarzschild
  %Black Hole,''
  Class.\ Quant.\ Grav.\  {\bf 25}, 175015 (2008)
  [arXiv:0801.4074 [gr-qc]];
  %%CITATION = CQGRD,25,175015;%%
%\cite{Nozari:2008gp}

\bibitem{others3}
  K.~Nozari and S.~Hamid Mehdipour,
  %``Quantum Gravity and Recovery of Information in Black Hole Evaporation,''
  Europhys.\ Lett.\  {\bf 84}, 20008 (2008)
  [arXiv:0804.4221 [gr-qc]];
  %%CITATION = EULEE,84,20008;%%
%\cite{Zhou:2008zzf}

\bibitem{others4}
  S.~Zhou and W.~Liu,
  %``Hawking radiation of charged Dirac particles from a Kerr-Newman black
  %hole,''
  Phys.\ Rev.\  D {\bf 77}, 104021 (2008);
  %%CITATION = PHRVA,D77,104021;%%
%\cite{Clifton:2008sb}

\bibitem{others5}
  T.~Clifton,
  %``Properties of Black Hole Radiation From Tunnelling,''
  Class.\ Quant.\ Grav.\  {\bf 25}, 175022 (2008)
  [arXiv:0804.2635 [gr-qc]];
  %%CITATION = CQGRD,25,175022;%%
%\cite{Liu:2006jw}

\bibitem{others6}
  C.~Z.~Liu, J.~Y.~Zhang and Z.~Zhao,
  %``Charged particle's tunneling from a dilaton black hole,''
  Phys.\ Lett.\  B {\bf 639}, 670 (2006);
  %%CITATION = PHLTA,B639,670;%%
%\cite{Liu:2008zzu}

\bibitem{others7}
  C.~Z.~Liu and Z.~Zhao,
  %``Hawking radiation via tunneling from general static black holes,''
  Mod.\ Phys.\ Lett.\  A {\bf 23}, 539 (2008);
  %%CITATION = MPLAE,A23,539;%%
%\cite{Fang:2005xf}

\bibitem{others8}
  H.~Z.~Fang, J.~Ren and Z.~Zhao,
  %``Particle tunnels from Garfinkle-Horne black hole and Horowitz-Strominger
  %black hole,''
  Int.\ J.\ Mod.\ Phys.\  D {\bf 14}, 1699 (2005);
  %%CITATION = IMPAE,D14,1699;%%
%\cite{Liu:2005hj}

\bibitem{others9}
  W.~Liu,
  %``New coordinates for BTZ black hole and Hawking radiation via  tunnelling,''
  Phys.\ Lett.\  B {\bf 634}, 541 (2006)
  [arXiv:gr-qc/0512099];
  %%CITATION = PHLTA,B634,541;%%
%\cite{Kim:2007ep}

\bibitem{others10}
  S.~P.~Kim,
  %``Hawking Radiation as Quantum Tunneling in Rindler Coordinate,''
  JHEP {\bf 0711}, 048 (2007)
  [arXiv:0710.0915 [hep-th]].
  %%CITATION = JHEPA,0711,048;%%


%\cite{Kerner:2006vu}
\bibitem{Kerner}
  R.~Kerner and R.~B.~Mann,
  %``Tunnelling, Temperature and Taub-NUT Black Holes,''
  Phys.\ Rev.\  D {\bf 73}, 104010 (2006)
  [arXiv:gr-qc/0603019].
  %%CITATION = PHRVA,D73,104010;%%

%\cite{Volovik:1999fc}
\bibitem{Volovik1}
  G.~E.~Volovik,
  %``Simulation of Panleve-Gullstrand black hole in thin 3He-A film,''
  Pisma Zh.\ Eksp.\ Teor.\ Fiz.\  {\bf 69}, 662 (1999)
  [JETP Lett.\  {\bf 69}, 705 (1999)]
  [arXiv:gr-qc/9901077];
  %%CITATION = JTPLA,69,705;%%
%\cite{Volovik:1999cn}

\bibitem{Volovik2}
  G.~E.~Volovik,
  %``He-3 and universe parallelism,''
  arXiv:cond-mat/9902171.
  %%CITATION = COND-MAT/9902171;%%


%\cite{Angheben:2005rm}
\bibitem{Angheben}
  M.~Angheben, M.~Nadalini, L.~Vanzo and S.~Zerbini,
  %``Hawking radiation as tunneling for extremal and rotating black holes,''
  JHEP {\bf 0505}, 014 (2005)
  [arXiv:hep-th/0503081].
  %%CITATION = JHEPA,0505,014;%%

%\cite{Srinivasan:1998ty}
\bibitem{Complex1}
  K.~Srinivasan and T.~Padmanabhan,
  %``Particle production and complex path analysis,''
  Phys.\ Rev.\  D {\bf 60}, 024007 (1999)
  [arXiv:gr-qc/9812028];
  %%CITATION = PHRVA,D60,024007;%%
%\cite{Shankaranarayanan:2000gb}

\bibitem{Complex2}
  S.~Shankaranarayanan, K.~Srinivasan and T.~Padmanabhan,
  %``Method of complex paths and general covariance of Hawking radiation,''
  Mod.\ Phys.\ Lett.\  A {\bf 16}, 571 (2001)
  [arXiv:gr-qc/0007022];
  %%CITATION = MPLAE,A16,571;%%
%\cite{Shankaranarayanan:2000qv}

\bibitem{Complex3}
  S.~Shankaranarayanan, T.~Padmanabhan and K.~Srinivasan,
  %``Hawking radiation in different coordinate settings: Complex paths
  %approach,''
  Class.\ Quant.\ Grav.\  {\bf 19}, 2671 (2002)
  [arXiv:gr-qc/0010042];
  %%CITATION = CQGRD,19,2671;%%
%\cite{Shankaranarayanan:2003ya}

\bibitem{Complex4}
  S.~Shankaranarayanan,
  %``Temperature and entropy of Schwarzschild-de Sitter space-time,''
  Phys.\ Rev.\  D {\bf 67}, 084026 (2003)
  [arXiv:gr-qc/0301090];
  %%CITATION = PHRVA,D67,084026;%%
%\cite{Vagenas:2001qw}

\bibitem{Complex5}
  E.~C.~Vagenas,
  %``Complex paths and covariance of Hawking radiation in 2D stringy black
  %holes,''
  Nuovo Cim.\  B {\bf 117}, 899 (2002)
  [arXiv:hep-th/0111047].
  %%CITATION = NUCIA,B117,899;%%

%\cite{Kraus:1994by}
\bibitem{Kraus1}
  P.~Kraus and F.~Wilczek,
  %``Self-Interaction Correction to Black Hole Radiance,''
  Nucl.\ Phys.\  B {\bf 433}, 403 (1995)
  [arXiv:gr-qc/9408003];
  %%CITATION = NUPHA,B433,403;%%
%\cite{KeskiVakkuri:1996xp}

\bibitem{Kraus2}
  E.~Keski-Vakkuri and P.~Kraus,
  %``Microcanonical D-branes and back reaction,''
  Nucl.\ Phys.\  B {\bf 491}, 249 (1997)
  [arXiv:hep-th/9610045];
  %%CITATION = NUPHA,B491,249;%%
%\cite{Kraus:1994fj}

\bibitem{Kraus3}
  P.~Kraus and F.~Wilczek,
  %``Effect Of Selfinteraction On Charged Black Hole Radiance,''
  Nucl.\ Phys.\  B {\bf 437}, 231 (1995)
  [arXiv:hep-th/9411219].
  %%CITATION = NUPHA,B437,231;%%

%\cite{Chowdhury}
\bibitem{Chowdhury}
  B.~D.~Chowdhury,
  %``Problems with Tunneling of Thin Shells from Black Holes,''
  Pramana {\bf 70}, 593 (2008)
  [Pramana {\bf 70}, 3 (2008)]
  [arXiv:hep-th/0605197];
  %%CITATION = PRAMC,70,3;%%

%\cite{Pilling}
\bibitem{Pilling}
  T.~Pilling,
  %``Tunneling derived from Black Hole Thermodynamics,''
  Phys.\ Lett.\  B {\bf 660}, 402 (2008)
  [arXiv:0709.1624 [gr-qc]].
  %%CITATION = PHLTA,B660,402;%%

%\cite{Akhmedova1}
\bibitem{Akhmedova1}
  V.~Akhmedova, T.~Pilling, A.~De Gill and D.~Singleton,
  %``Temporal contribution to gravitational WKB-like calculations,''
  Phys.\ Lett.\  B {\bf 666}, 269 (2008)
  [arXiv:0804.2289 [hep-th]];
  %%CITATION = PHLTA,B666,269;%%
%\cite{Akhmedov2}

\bibitem{Akhmedova2}
  E.~T.~Akhmedov, V.~Akhmedova and D.~Singleton,
  %``Hawking temperature in the tunneling picture,''
  Phys.\ Lett.\  B {\bf 642} (2006) 124
  [arXiv:hep-th/0608098].
  %%CITATION = PHLTA,B642,124;%%

%\cite{Hawking}
\bibitem{Hawking}
  S.~W.~Hawking,
  %``Particle Creation By Black Holes,''
  Commun.\ Math.\ Phys.\  {\bf 43}, 199 (1975)
  [Erratum-ibid.\  {\bf 46}, 206 (1976)].
  %%CITATION = CMPHA,43,199;%%

%\cite{Gibbons:1976ue}
\bibitem{Euclidean}
  G.~W.~Gibbons and S.~W.~Hawking,
  %``Action Integrals And Partition Functions In Quantum Gravity,''
  Phys.\ Rev.\  D {\bf 15}, 2752 (1977).
  %%CITATION = PHRVA,D15,2752;%%

%\cite{Damour:1976jd}
\bibitem{D-R method}
  T.~Damour and R.~Ruffini,
  %``Black Hole Evaporation In The Klein-Sauter-Heisenberg-Euler Formalism,''
  Phys.\ Rev.\  D {\bf 14}, 332 (1976).
  %%CITATION = PHRVA,D14,332;%%

%\cite{Robinson:2005pd}
\bibitem{Anomaly1}
  S.~P.~Robinson and F.~Wilczek,
  %``A relationship between Hawking radiation and gravitational anomalies,''
  Phys.\ Rev.\ Lett.\  {\bf 95}, 011303 (2005)
  [arXiv:gr-qc/0502074];
  %%CITATION = PRLTA,95,011303;%%
%\cite{Iso:2006wa}

\bibitem{Anomaly2}
  S.~Iso, H.~Umetsu and F.~Wilczek,
  %``Hawking radiation from charged black holes via gauge and gravitational
  %anomalies,''
  Phys.\ Rev.\ Lett.\  {\bf 96}, 151302 (2006)
  [arXiv:hep-th/0602146];
  %%CITATION = PRLTA,96,151302;%%
%\cite{Iso:2006ut}

\bibitem{Anomaly3}
  S.~Iso, H.~Umetsu and F.~Wilczek,
  %``Anomalies, Hawking radiations and regularity in rotating black holes,''
  Phys.\ Rev.\  D {\bf 74}, 044017 (2006)
  [arXiv:hep-th/0606018];
  %%CITATION = PHRVA,D74,044017;%%
%\cite{Vagenas:2006qb}

\bibitem{Anomaly4}
  E.~C.~Vagenas and S.~Das,
  %``Gravitational anomalies, Hawking radiation, and spherically symmetric
  %black holes,''
  JHEP {\bf 0610}, 025 (2006)
  [arXiv:hep-th/0606077];
  %%CITATION = JHEPA,0610,025;%%
%\cite{Das:2007ru}

\bibitem{Anomaly5}
  S.~Das, S.~P.~Robinson and E.~C.~Vagenas,
  %``Gravitational anomalies: a recipe for Hawking radiation,''
  Int.\ J.\ Mod.\ Phys.\  D {\bf 17}, 533 (2008)
  [arXiv:0705.2233 [hep-th]];
  %%CITATION = IMPAE,D17,533;%%
%\cite{Banerjee:2008wq}

\bibitem{Anomaly6}
  R.~Banerjee and S.~Kulkarni,
  %``Hawking Radiation, Covariant Boundary Conditions and Vacuum States,''
  arXiv:0810.5683 [hep-th];
  %%CITATION = ARXIV:0810.5683;%%
%\cite{Banerjee:2007uc}

\bibitem{Anomaly7}
  R.~Banerjee and S.~Kulkarni,
  %``Hawking Radiation, Effective Actions and Covariant Boundary Conditions,''
  Phys.\ Lett.\  B {\bf 659}, 827 (2008)
  [arXiv:0709.3916 [hep-th]];
  %%CITATION = PHLTA,B659,827;%%
%\cite{Banerjee:2007qs}

\bibitem{Anomaly8}
  R.~Banerjee and S.~Kulkarni,
  %``Hawking Radiation and Covariant Anomalies,''
  Phys.\ Rev.\  D {\bf 77}, 024018 (2008)
  [arXiv:0707.2449 [hep-th]].
  %%CITATION = PHRVA,D77,024018;%%

%\cite{Mitra:2006qa}
\bibitem{Mitra1}
  P.~Mitra,
  %``Hawking temperature from tunnelling formalism,''
  Phys.\ Lett.\  B {\bf 648}, 240 (2007)
  [arXiv:hep-th/0611265];
  %%CITATION = PHLTA,B648,240;%%
%\cite{Stotyn:2008qu}

\bibitem{Mitra2}
  S.~Stotyn, K.~Schleich and D.~Witt,
  %``Observer Dependent Horizon Temperatures: a Coordinate-Free Formulation of
  %Hawking Radiation as Tunneling,''
  arXiv:0809.5093 [gr-qc].
  %%CITATION = ARXIV:0809.5093;%%

\bibitem{relation}
  J.~Y.~Zhang, Y.~P.~Hu and Z.~Zhao,
  %``Information loss in black hole evaporation,''
  Mod.\ Phys.\ Lett.\  A {\bf 21}, 1865 (2006)
  [arXiv:hep-th/0512121];
  %%CITATION = MPLAE,A21,1865;%%

%\cite{Zhang:2009ym}
\bibitem{Baocheng Zhang}
  B.~Zhang, Q.~Y.~Cai and M.~S.~Zhan,
  %``The temperature in Hawking radiation as tunneling,''
  Phys.\ Lett.\  B {\bf 671}, 310 (2009)
  [arXiv:0901.0591 [hep-th]].
  %%CITATION = PHLTA,B671,310;%%

%\cite{Bekenstein:1973ur}
\bibitem{Bekenstein1}
  J.~D.~Bekenstein,
  %``Black holes and entropy,''
  Phys.\ Rev.\  D {\bf 7}, 2333 (1973);
  %%CITATION = PHRVA,D7,2333;%%
%\cite{Bekenstein:1974ax}

\bibitem{Bekenstein2}
  J.~D.~Bekenstein,
  %``Generalized second law of thermodynamics in black hole physics,''
  Phys.\ Rev.\  D {\bf 9}, 3292 (1974).
  %%CITATION = PHRVA,D9,3292;%%

\bibitem{DTherml1}
S.~A.~Hayward, %`` General laws of black-hole dynamics%
Phys.\ Rev.\ D \textbf{49}, 6467 (1994);

\bibitem{DTherml2} S.~A.~Hayward,
%``Gravitational energy in spherical symmetry,''
Phys.\ Rev.\ D \textbf{53}, 1938 (1996) [arXiv:gr-qc/9408002];

\bibitem{DTherml3} S.~A.~Hayward,
%``Unified first law of black-hole dynamics and relativistic thermodynamics,''
Class.\ Quant.\ Grav.\ \textbf{15}, 3147 (1998)
[arXiv:gr-qc/9710089];

\bibitem{DTherml4} S.~A.~Hayward, S.~Mukohyama and M.~C.~Ashworth,
%``Dynamic black-hole entropy,''
Phys.\ Lett.\ A \textbf{256}, 347 (1999) [arXiv:gr-qc/9810006];

\bibitem{DTherml5} R.~G.~Cai and L.~M.~Cao,
%``Unified first law and thermodynamics of apparent horizon in FRW
%universe,''
Phys.\ Rev.\ D \textbf{75}, 064008 (2007) [arXiv:gr-qc/0611071];
%%CITATION = PHRVA,D75,064008;%%
%\cite{Akbar:2008vz}

\bibitem{DTherml6}
  M.~Akbar,
  %``Generalized Second Law of Thermodynamics in Extended Theories of Gravity,''
  arXiv:0808.3308 [gr-qc];
  %%CITATION = ARXIV:0808.3308;%%
%\cite{Akbar:2008vc}

\bibitem{DTherml7}
  M.~Akbar,
  %``Viscous Cosmology and Thermodynamics of Apparent Horizon,''
  arXiv:0808.0169 [gr-qc];
  %%CITATION = ARXIV:0808.0169;%%
%\cite{Cai:2008ys}

\bibitem{DTherml8}
  R.~G.~Cai, L.~M.~Cao and Y.~P.~Hu,
  %``Corrected Entropy-Area Relation and Modified Friedmann Equations,''
  JHEP {\bf 0808}, 090 (2008)
  [arXiv:0807.1232 [hep-th]];
  %%CITATION = JHEPA,0808,090;%%
%\cite{Wu:2008rp}

\bibitem{DTherml9}
  S.~F.~Wu, G.~H.~Yang and P.~M.~Zhang,
  %``The equation of state for scalar-tensor gravity,''
  arXiv:0805.4044 [hep-th];
  %%CITATION = ARXIV:0805.4044;%%
%\cite{Zhu:2008di}

\bibitem{DTherml10}
  T.~Zhu, J.~R.~Ren and S.~F.~Mo,
  %``Thermodynamics of Friedmann Equation and Masslike Function in Generalized
  %Braneworlds,''
  arXiv:0805.1162 [gr-qc];
  %%CITATION = ARXIV:0805.1162;%%
%\cite{Wu:2008ir}

\bibitem{DTherml11}
  S.~F.~Wu, B.~Wang, G.~H.~Yang and P.~M.~Zhang,
  %``The generalized second law of thermodynamics in generalized gravity
  %theories,''
  arXiv:0801.2688 [hep-th];
  %%CITATION = ARXIV:0801.2688;%%
%\cite{Cai:2007bh}

\bibitem{DTherml12}
  R.~G.~Cai,
  %``Thermodynamics of Apparent Horizon in Brane World Scenarios,''
  Prog.\ Theor.\ Phys.\ Suppl.\  {\bf 172}, 100 (2008)
  [arXiv:0712.2142 [hep-th]];
  %%CITATION = PTPSA,172,100;%%
%\cite{Wu:2007se}

\bibitem{DTherml13}
  S.~F.~Wu, B.~Wang and G.~H.~Yang,
  %``Thermodynamics on the apparent horizon in generalized gravity theories,''
  Nucl.\ Phys.\  B {\bf 799}, 330 (2008)
  [arXiv:0711.1209 [hep-th]];
  %%CITATION = NUPHA,B799,330;%%
%\cite{Wu:2007em}

\bibitem{DTherml14}
  S.~F.~Wu, G.~H.~Yang and P.~M.~Zhang,
  %``Cosmological equations and Thermodynamics on Apparent Horizon in Thick
  %Braneworld,''
  arXiv:0710.5394 [hep-th];
  %%CITATION = ARXIV:0710.5394;%%
%\cite{Zhou:2007pz}

\bibitem{DTherml15}
  J.~Zhou, B.~Wang, Y.~Gong and E.~Abdalla,
  %``The second law of thermodynamics in the accelerating universe,''
  Phys.\ Lett.\  B {\bf 652}, 86 (2007)
  [arXiv:0705.1264 [gr-qc]];
  %%CITATION = PHLTA,B652,86;%%

\bibitem{DTherml16}
  J.~R.~Ren and R.~Li,
  %``Unified First Law and Thermodynamics of Dynamical Black Hole in
  %n-dimensional Vaidya Spacetime,''
  arXiv:0705.4339 [gr-qc];
  %%CITATION = ARXIV:0705.4339;%%
%\cite{Gong:2007md}

\bibitem{DTherml17}
  Y.~Gong and A.~Wang,
  %``The Friedmann equations and thermodynamics of apparent horizons,''
  Phys.\ Rev.\ Lett.\  {\bf 99}, 211301 (2007)
  [arXiv:0704.0793 [hep-th]];
  %%CITATION = PRLTA,99,211301;%%
%\cite{Sheykhi:2007gi}

\bibitem{DTherml18}
  A.~Sheykhi, B.~Wang and R.~G.~Cai,
  %``Deep Connection Between Thermodynamics and Gravity in Gauss-Bonnet
  %Braneworld,''
  Phys.\ Rev.\  D {\bf 76}, 023515 (2007)
  [arXiv:hep-th/0701261];
  %%CITATION = PHRVA,D76,023515;%%
%\cite{Sheykhi:2007zp}

\bibitem{DTherml19}
  A.~Sheykhi, B.~Wang and R.~G.~Cai,
  %``Thermodynamical Properties of Apparent Horizon in Warped DGP Braneworld,''
  Nucl.\ Phys.\  B {\bf 779}, 1 (2007)
  [arXiv:hep-th/0701198];
  %%CITATION = NUPHA,B779,1;%%
%\cite{Padmanabhan:2007en}

\bibitem{DTherml20}
  T.~Padmanabhan and A.~Paranjape,
  %``Entropy of Null Surfaces and Dynamics of Spacetime,''
  Phys.\ Rev.\  D {\bf 75}, 064004 (2007)
  [arXiv:gr-qc/0701003];
  %%CITATION = PHRVA,D75,064004;%%
%\cite{Cai:2006pa}

\bibitem{DTherml21}
  R.~G.~Cai and L.~M.~Cao,
  %``Thermodynamics of Apparent Horizon in Brane World Scenario,''
  Nucl.\ Phys.\  B {\bf 785}, 135 (2007)
  [arXiv:hep-th/0612144];
  %%CITATION = NUPHA,B785,135;%%
%\cite{Akbar:2006mq}

\bibitem{DTherml22}
  M.~Akbar and R.~G.~Cai,
  %``Thermodynamic Behavior of Field Equations for f(R) Gravity,''
  Phys.\ Lett.\  B {\bf 648}, 243 (2007)
  [arXiv:gr-qc/0612089];
  %%CITATION = PHLTA,B648,243;%%
%\cite{Akbar:2006kj}

\bibitem{DTherml23}
  M.~Akbar and R.~G.~Cai,
  %``Thermodynamic Behavior of Friedmann Equation at Apparent Horizon of FRW
  %Universe,''
  Phys.\ Rev.\  D {\bf 75}, 084003 (2007)
  [arXiv:hep-th/0609128];
  %%CITATION = PHRVA,D75,084003;%%
%\cite{Akbar:2006er}

\bibitem{DTherml24}
  M.~Akbar and R.~G.~Cai,
  %``Friedmann equations of FRW universe in scalar-tensor gravity, f(R)  gravity
  %and first law of thermodynamics,''
  Phys.\ Lett.\  B {\bf 635}, 7 (2006)
  [arXiv:hep-th/0602156];
  %%CITATION = PHLTA,B635,7;%%

\bibitem{DTherml25}
R.~G.~Cai and S.~P.~Kim, JHEP {\bf 0502}, 050 (2005)
[arXiv:hep-th/0501055];
%\cite{Cai:2008mh}

\bibitem{DTherml26}
  R.~G.~Cai, L.~M.~Cao, Y.~P.~Hu and S.~P.~Kim,
  %``Generalized Vaidya Spacetime in Lovelock Gravity and Thermodynamics on
  %Apparent Horizon,''
  Phys.\ Rev.\  D {\bf 78}, 124012 (2008)
  [arXiv:0810.2610 [hep-th]].
  %%CITATION = PHRVA,D78,124012;%%

%\cite{Sannan:1988eh}
\bibitem{Sannan}
  S.~Sannan,
  %``HEURISTIC DERIVATION OF THE PROBABILITY DISTRIBUTIONS OF PARTICLES EMITTED
  %BY A BLACK HOLE,''
  Gen.\ Rel.\ Grav.\  {\bf 20}, 239 (1988).
  %%CITATION = GRGVA,20,239;%%



\end{thebibliography}
\end{document}